\documentclass{ws-procs9x6}

\usepackage{graphicx}
\usepackage{color}

\definecolor{mycolor}{rgb}{0.9, 0.2, 0.7}
\definecolor{sbc}{rgb}{0.1, 0.9, 0.1}

\begin{document}

\title{CLOUDS, CLUMPS, CORES \& COMETS - \\  A COSMIC CHEMICAL CONNECTION ?  }


\author{S.B. CHARNLEY  \&  S.D. RODGERS }

\address{Space Science \& Astrobiology Division, MS 245-3, NASA Ames
 Research Center \\
 Moffett Field, CA 94035-1000, USA \\
 E-mail: charnley@dusty.arc.nasa.gov}


\begin{abstract}

We discuss  the connection between the chemistry of dense interstellar clouds  and those characteristics of cometary matter that could be remnants of it. The chemical evolution observed to occur in molecular clouds is summarized and a model for dense core collapse that can plausibly account for  the isotopic fractionation of hydrogen, nitrogen, oxygen and carbon measured in primitive solar system materials is presented.

\end{abstract}


\bodymatter

\section{Introduction}
\label{sec:intro}

Comets have probably retained some material that originated in the
molecular cloud from which the Sun formed\cite{Pascale04}. Determining
how much pristine interstellar material is in comets will help answer
many important questions relating to the origin of our Solar System
and, as comets are strong candidates for seeding planets with complex
organic molecules, understanding the details of the interstellar-comet
connection will have important implications for
astrobiology\cite{ARAA00}. Recent data from the {\it Stardust}
mission\cite{Brownlee06}, and ground-based observations\cite{Wooden02,Wooden07},
indicate that some materials present in cometary dust experienced very
high temperatures ($\sim 800$ K), relative to those typically found in
molecular clouds ($\sim 10$ K).  Nevertheless, the organic inventory
and isotopic signatures measured for cometary molecules do provide a
tantalizing connection with interstellar chemistry.
 
In this paper we discuss the chemical structure and evolution of
interstellar matter prior to its incorporation into protoplanetary
disks. We outline a model whereby many of the key cosmogonic markers
in cometary matter can be explained as being of interstellar origin.


\section{Molecular clouds and star formation}
\label{sec:sf}

The Sun, the planets, and other small solar system bodies, originate
from the protostar and gaseous dusty disk produced by the
gravitational collapse of molecular cloud material\cite{PPVBook}{}.
Prior to incorporation into a protostellar/protoplanetary
disk, the chemistry of dense interstellar matter can be tracked by
monitoring the evolutionary state of the cloud and embedded
protostar(s).   We can identify 5 stages of direct interest.  First, there
is the {\it ambient dense cloud material}. This is found to be clumpy
and turbulent, with hydrogen densities of $ \sim 10^3-10^4 \rm
cm^{-3}$ and kinetic temperatures of $\sim 10$K.  In dense clouds,
{\it prestellar (or starless) cores} are apparently globular
structures with pronounced density gradient towards the center ($ >
10^5 \rm cm^{-3}$)\cite{Lada07}, and are where low-mass
protostars can eventually form\cite{BerginARAA07}{}. Protostars
at the earliest stage of evolution, the {\it Class 0} sources, are
deeply embedded in their parent core from which they, and their
associated disks, accrete mass at a very high rate.  As sources
undergo transition into the {\it Class I } epoch, the optically
invisible protostar has accumulated most of its mass from the
circumstellar envelope and a gaseous disk is in Keplerian orbit around
it.  Objects at the {\it Class II} stage are optically visible T Tauri
stars and most of their circumstellar gas is contained in the disk.

\begin{figure*}    
\begin{center}
\includegraphics[scale=0.7]{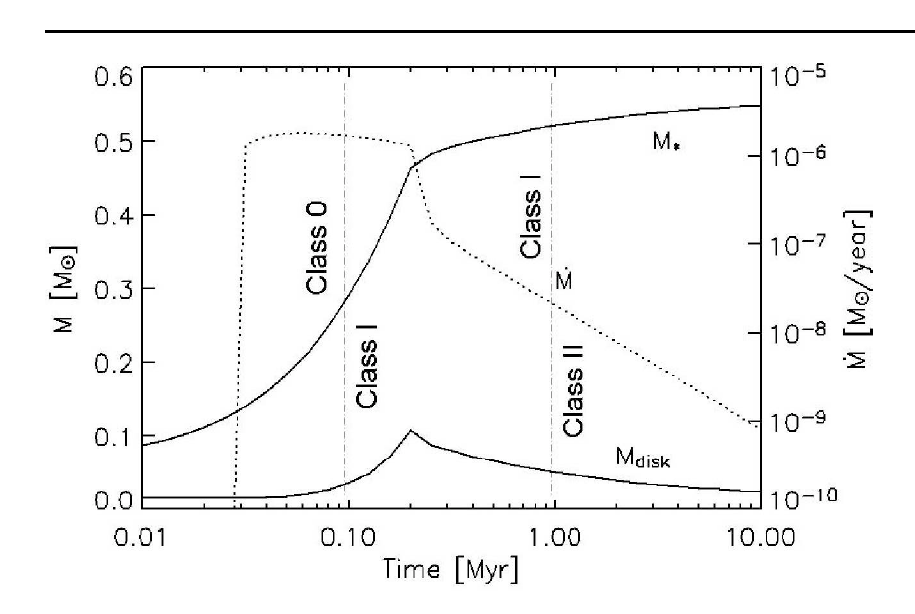}  
\caption{Evolution of protostellar and disk masses\cite{vanD07,Hueso05}.
}
\label{fig:hueso}
\end{center}  
\end{figure*}

  Figure 1 illustrates the time-scales associated with
protostar-disk mass accumulation (see Ref.~\refcite{Hueso05}). 
Detailed observational studies and chemical modelling have led to a deeper understanding of the  
chemistry of the evolutionary sequence from pre-collapse (and prestellar)
cores, through to Class 0 and I sources,   Class II and III objects
and their related `protoplanetary' and `debris' disks
 \cite{Thi04,vanD06,diFrancescoPPV,CeccPPV,BerginPPV,
vanDBlake98,Mundy00,Evans06,Ceccarelli06,Lada07,SDRCocoon,Lee05}
Such studies can  also shed light on the state of the interstellar material
available to disks during the comet-formation epoch  (e.g.\
Ref.~\refcite{AJM04}).  
%
%
As most of the mass accreted up to early in the Class I epoch
was consumed by growth of the protosun (Figure 1), the volatile disk material now
retained in comets probably accreted during the Class I-Class II
evolutionary phases, where the lower mass accretion rates also favour much
weaker accretion shock strengths over the disk surface\cite{Cassen81}.
Thus, the chemistry in molecular clouds containing
protostars at the Class 0/I boundary, and later, is that which may be
best associated with the most pristine cometary matter.

 
\begin{figure*}[t]    
\begin{center} 
\includegraphics[scale=0.4]{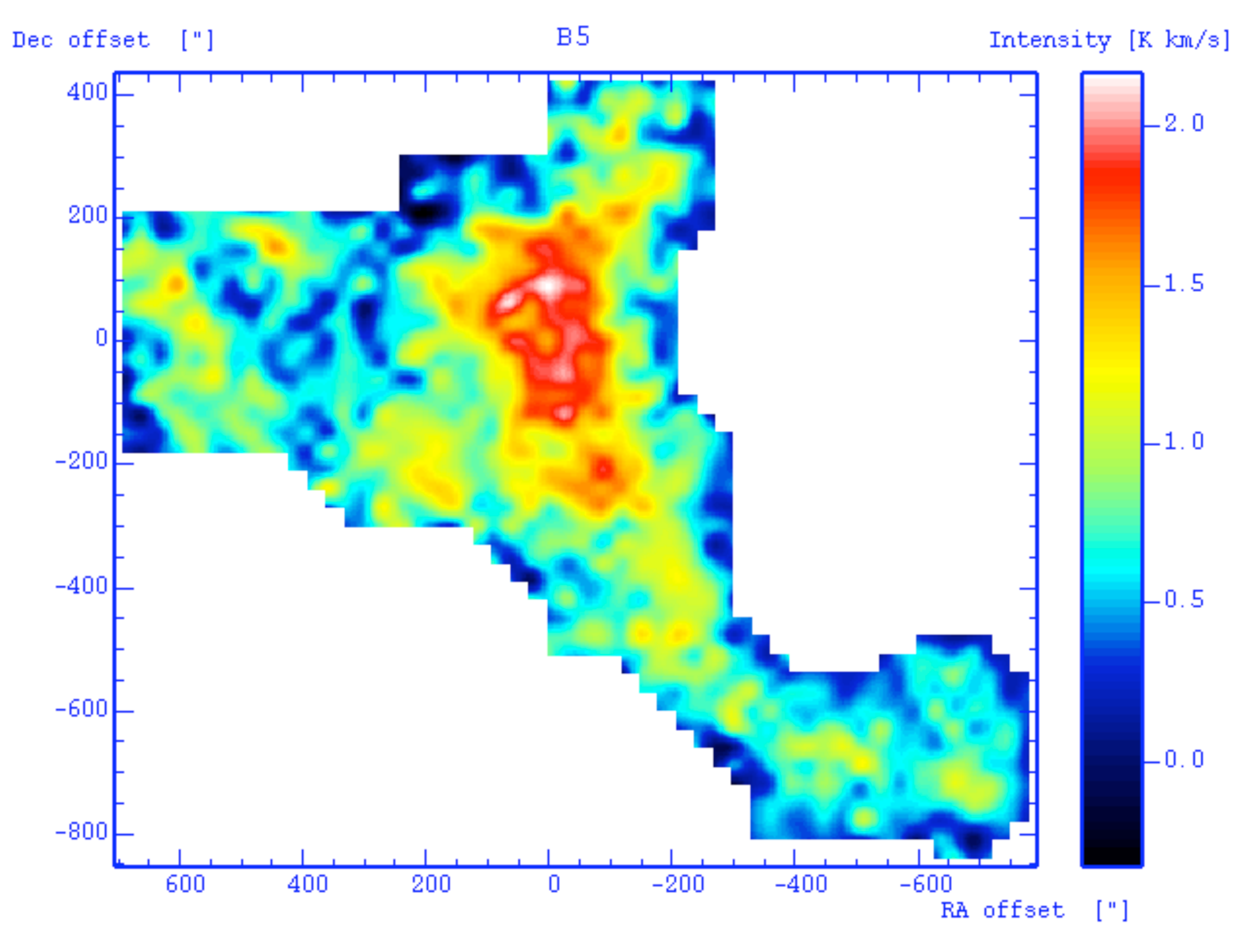}
\caption{Onsala 20m telescope map of C$^{18}$O  $J$=1-0 emission from the Barnard 5
 cloud \protect{\cite{Butner07}}.}
\label{fig:B5CO}
\end{center}  
\end{figure*} 
  
\subsection{Observations of cloud chemical evolution} 
\label{b5}
       
The Barnard 5 molecular cloud in Perseus (B5) is a region in which
low-mass protostars are forming\cite{Goldsmith86,Fuller91}{}.
Of the four identified protostars\cite{Beichman84}, B5 IRS 1 resides
in the dense central core and is classified as a Class I source that
may just be at the Class 0/I transition\cite{Yu99}{}. Interferometric
maps of IRS1 confirm the presence of a circumstellar disk of $\sim
0.16-0.27$ solar masses \cite{Langer96}{}.  Mapping of clouds like B5
in various molecular lines\cite{Buckle06,diFrancescoPPV,BerginARAA07}
allows us to understand the physical and chemical evolution that
occurs in cloud material prior to incorporation into a circumstellar
disk.



\begin{figure*}[t]    
\begin{center} 
\hbox{\hspace{2.0cm}
 \includegraphics[angle=0,scale=0.20]{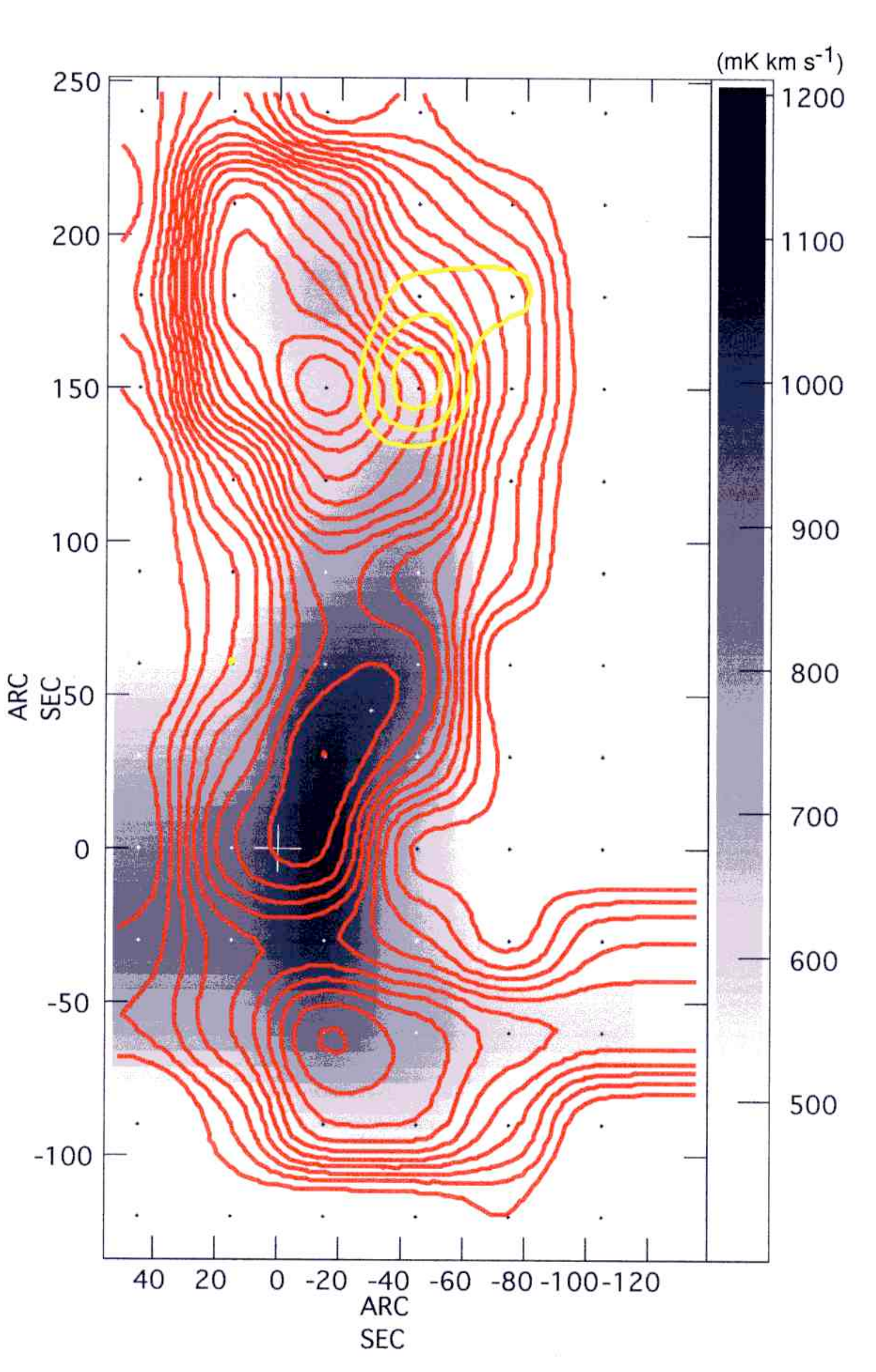} 
 \includegraphics[angle=0,scale=0.20]{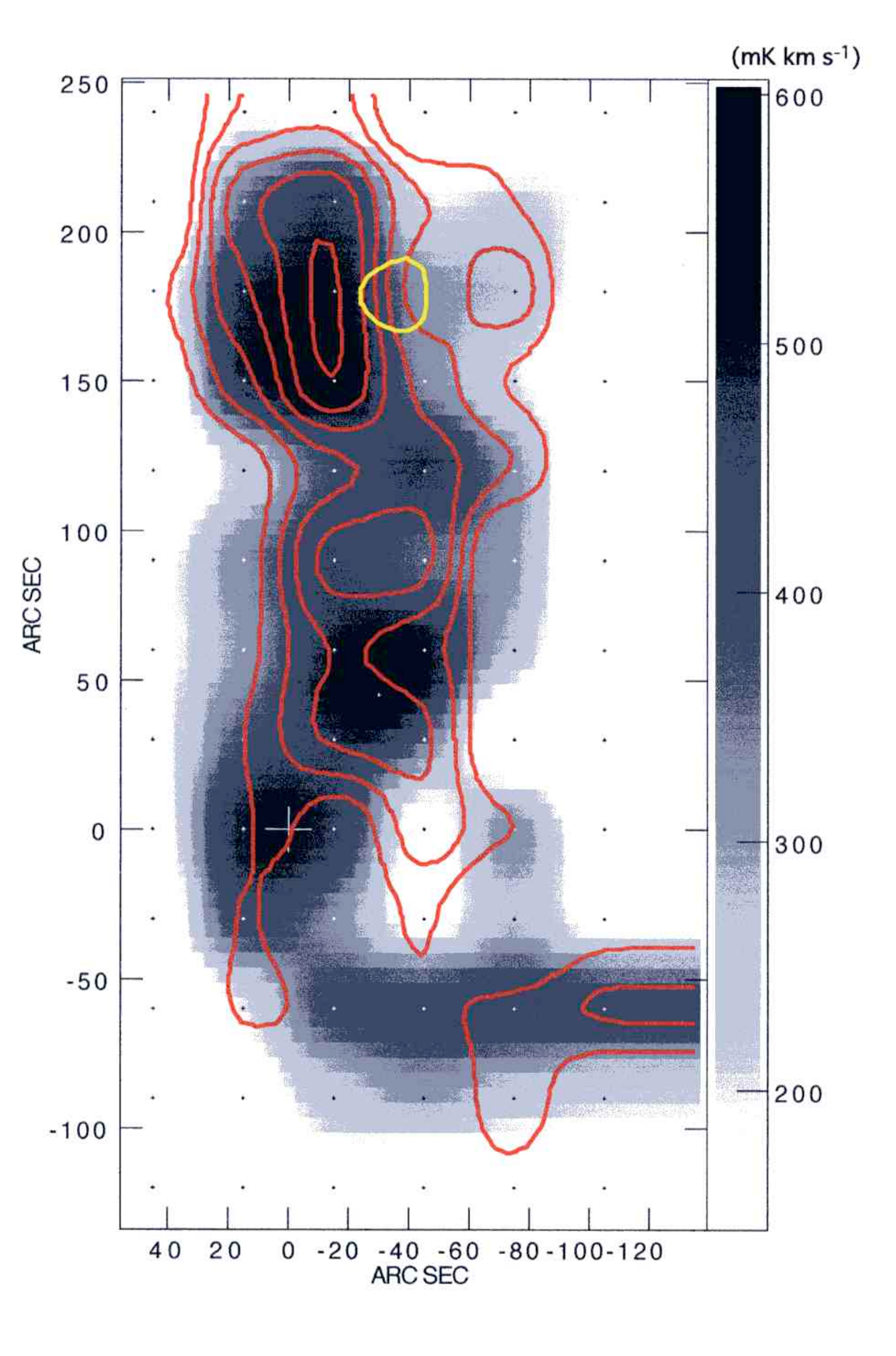}   
}
\caption{Nobeyama 45m telescope integrated intensity maps of the
  central region of B5\cite{Charnleyetal07}.
  The (0,0) position is B5 IRS1 which is marked by a
  small cross.  The left panel shows the NH$_3$ (1,1) map in grey
  scale. Red and yellow contours respectively mark the HC$_3$N($J$=5-4) and HC$_5$N$(J$=17-16) maps,.  In the right panel, the $c$-C$_3$H$_2$(2$_{12}$-1$_{01}$) integrated intensity is the grey scale,   the CCS(4$_3$-3$_2$) map  is shown as  red
  contours , and the yellow contour is the C$_3$S($J$=4-3) emission  peak. }
\label{fig:shige}
\end{center}  
\end{figure*}

High resolution C$^{18}$O mapping of the B5 core (Figure
  2) shows a clumpy morphology for both the extended ambient gas and
the dense central core\cite{Butner07}.  Maps of the B5 core
(Figure 3) demonstrate that other molecular
distributions are also very clumpy and chemically
differentiated\cite{Charnleyetal07}.  This chemical anticorrelation between the
emission peaks of many molecules is evident in other high-resolution
maps of the B5 core (e.g., CH$_3$OH, H$^{13}$CO$^+$)\cite{Butner07},
and other dark clouds \cite{Buckle06}{}. In particular,
 Figure 3  shows that emission from various carbon-chain
species is anti-correlated with that of ammonia, as observed in TMC-1
and several other dense cores\cite{Olano88,Hirahara92,BerginARAA07}
 Chemical models indicate that
carbon-chain species such as C$_2$S, C$_4$H, and the cyanopolyynes
(HC$_{2n+1}$N, $n$ =1-5) are `early time' species - they reach their
peak abundances in $\sim 10^5$ years - whereas `late time' species,
including N$_2$H$^+$ and NH$_3$, peak much later\cite{HerbstLeung89}{}.
This has led to the idea that observed spatial compositional
gradients (e.g.\ involving the carbon-chains and ammonia) are due to
differing chemical ages within individual clouds\cite{Hirahara92,
Peng98,Ohishi98,AJM2000,BerginARAA07}.
Thus, the two C-chain peaks (CP1\&2), near $\delta \sim +150''$, are
probably at an earlier stage of evolution relative to the material
surrounding IRS1, as one would expect.

In B5 we can observe the chemistry at three distinct phases in the
star formation sequence: the clumpy ambient medium, dense prestellar
clumps and a core containing a Class I protostar.  However, in B5 we
do not appear to see the chemistry at two intervening phases.
   

\begin{figure*}[t]    
\begin{center} 
\includegraphics[scale=2.0]{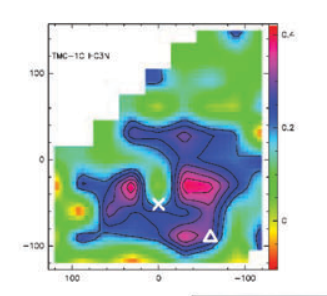}  
\caption{Integrated intensity map of the HC$_3$N $J$=10-9 emission in
  the TMC-1C cloud (K km/sec levels indicated by the scale to the
  right).  The peak in the N$_2$H$^+$ emission found by Caselli {\it et
  al.}\cite{Caselli02} is indicated by a cross. The approximate peak in the
  diffuse submillimetre emission is indicated by a triangle.  From
  Ref.~\refcite{Buckle06}.}
\label{fig:buckle}
\end{center}  
\end{figure*}

Maps of starless globules in lines of carbon chain molecules show that
they can have a reasonably smooth emission\cite{Hirota02,Hirota04}.
An apparently common subsequent step from such dense
prestellar clumps is the formation of a {\it depletion core} where
observations indicate that almost all the heavy molecules,
particularly CO, become depleted into the solid state\cite{Bergin01,
Bergin02,Tafalla02},
and strong emission from
deuterated molecules becomes evident\cite{CeccPPV}.
Hence, all the elemental C, O, N and S can be available for
grain-surface chemistry.  For example,  Figure 4  shows
the dense core of the starless dark globule TMC-1C mapped in its
HC$_3$N emission\cite{Buckle06}. This morphology can be
explained as resulting from the increasingly efficient sticking of
HC$_3$N molecules on dust as the density increases, the end-point
being a core with all the HC$_3$N molecules frozen out in the center.
Comparison with an N$_2$H$^+$ map of this source\cite{Caselli02}
indicates that the N$_2$H$^+$ emission actually peaks at the central
`hole' in the HC$_3$N emission (see  Figure 4).  This is
an example of {\it selective depletion} where the enhancement of
N-bearing species (such as N$_2$H$^+$ and N$_2$D$^+$) in CO-poor
regions is thought to be related to the relative depletion of CO and
N$_2$ \cite{SBC94,SBC97Cores,Bergin97}
 
Although depletion cores are very difficult to study both in molecular
lines (except perhaps for N$_2$H$^+$, N$_2$D$^+$, H$_2$D$^+$ and HD$_2 ^+$) and also in solid
state absorption, the chemistry occurring in their ices can be
revealed later in the star formation sequence.  Many cores at the
prestellar stage show evidence for gravitational
infall\cite{diFrancescoPPV} 
and the next step should be formation of a protostar in the deeply
embedded Class 0 phase\footnote{A candidate Class 0 source has
  recently been identified in B5 (J.V. Buckle, priv. comm.) that would
  be coincident with a putative depletion core identified in CH$_3$OH
  maps\protect{\cite{Butner07}}}. During the Class 0 phase, the protostar
and disk accrete mass at a high rate; the consequent increase in
luminosity and the present of outflows and shock waves means that it
heats its immediate environment and a {\it hot corino} is
formed\cite{CeccPPV}. 
Observations of hot corinos show that, like their massive
counterparts, the hot molecular cores\cite{ARAA00} 
they are very rich in complex organic molecules and exhibit extremely
pronounced deuterium fractionation characteristics\cite{CeccPPV}.
This is believed to be due to the evaporation and/or
sputtering of molecular ices into the gas\cite{Wooden04}.
Well-studied `hot corinos' are IRAS 16293-2422, NGC1333-IRAS4A,
NGC1333-IRAS4B and NGC1333-IRAS2A\cite{Cazeaux03,Kuan04,Bottinelli04a,
Bottinelli04b,Jorgensen05,Sakai06,Bottinelli07}.

Hence, the protostellar disk of a source such as B5 IRS 1 may, in
principle, accrete material with chemical characteristics
representative of a prestellar core, a depletion core, the Class 0 hot
corino, and the remnant envelope at the Class I phase. The actual
relative proportion of each chemistry that could end up unaltered in
comets is unknown.  Comparison with the measured volatile composition
of comets can help us better understand this putative connection and
to develop and constrain chemical models of the passage of
interstellar material into disks.


\section{Measured characteristics of cometary material}
\label{sbc:sec:comets}

The inventory of molecules observed in comets is, in general, a subset
of the $\sim 150$ molecules detected in the ISM\@. 
\Tref{sbc:tab:molobs} lists  species detected in interstellar
and circumstellar environments, with
those also detected in comets highlighted. The extensive overlap
between cometary and interstellar molecules has long suggested a
connection between the two\cite{Mayo}\@. A thorough description of all
cometary characteristics and their possible links with the ISM is
beyond the scope of this paper, and has been discussed in detail
elsewhere \cite{IrvPPIV,ARAA00,Pascale04,ISSI}\@.
In the following discussion, we focus on isotopic ratios in specific molecules,
and their potential role as tracers of the cometary--ISM connection.

\begin{table}    
\tbl{Interstellar, circumstellar, and cometary molecules}
{
\includegraphics[width=4.5in]{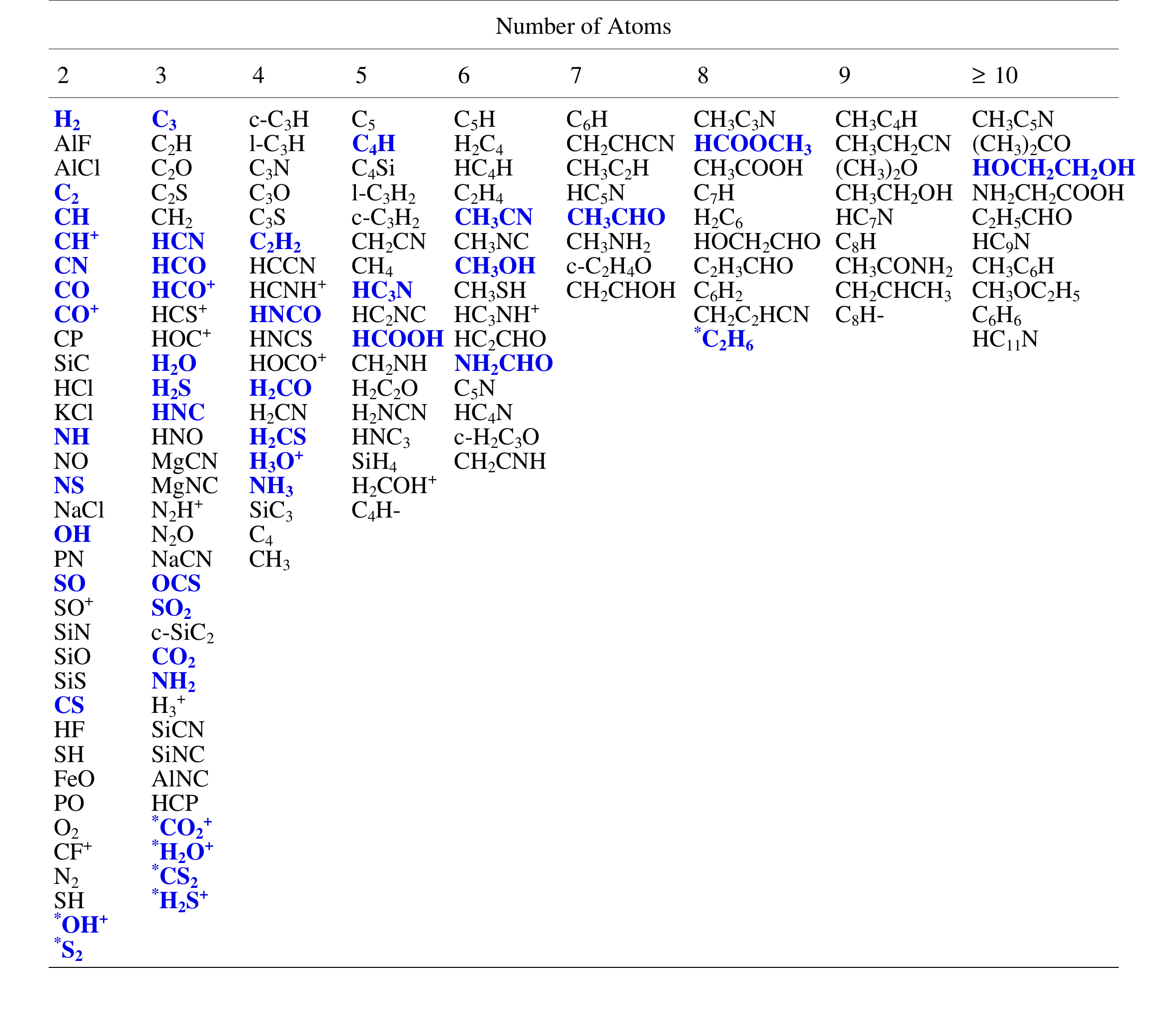}  
}
\begin{tabnote}
{\small Molecules in  blue are detected in
 comets. Those marked with an asterisk are uniquely detected in
 comets. Data from \\ {\tt
   www.cv.nrao.edu/$\sim$awootten/allmols.html}
 and {\tt www.astrochymist.org } }
\end{tabnote}
\label{sbc:tab:molobs}
\end{table}


\subsection{Hydrogen and deuterium}

Cometary D/H ratios are not significantly altered in the
coma\cite{RC02}, and are not thought to be altered during sublimation,
or during their long deep-freeze storage over the age of the Solar
System. Enhanced D/H ratios are a hallmark of interstellar
material\cite{Roueff03}, and the presence of highly-fractionated
material in comets would be strong evidence for the preservation of
interstellar material. D/H ratios have been observed in a handful of
molecules in a total of five different comets, and are summarized in
\tref{sbc:tab:dh}. In contrast, large numbers of deuterated species
have been observed in the ISM, including several multiply-deuterated
species\cite{CeccPPV}\@. 
Deuterium enrichments have also been seen in cometary dust grains
returned by the {\it Stardust} mission\cite{Sand06}, and in
interplanetary dust particles thought to originate from
comets\cite{Mess00}\@. This D-enrichment is also thought to originate
in the ISM, but as it resides in complex organic refractory material
it is not possible to trace a direct link to specific interstellar
molecules. 

\begin{table}[t]
\tbl{D/H ratios in comets and the ISM}
{
\begin{tabular}{@{}rlllll@{}}
\toprule 
Molecule  &  Cometary     & Comet    & Interstellar$^a$ & Refs.$^a$ \\
          &  D/H Ratio    &          &      D/H Ratio   & \\
\colrule
HDO/H$_2$O  &     ~~~~0.0003   & Several$^b$  &  0.0004 -- 0.01 & $^b$ \\
DCN/HCN     &    ~~~~0.002   &Hale-Bopp    &   0.01 -- 0.1  & \protect{\refcite{Meier98b}} \\
HDCO/H$_2$CO     & ~~~0.28 & T7 (LINEAR) &   0.07 -- 0.3 & \protect{\refcite{Kuan07}} \\
CH$_3$D/CH$_4$ &  $<0.04$ &  Q4 (NEAT)  & $<$0.06  & \protect{\refcite{Kawa05}} \\
 NH$_2$D/NH$_3$  & $<0.04$  & Hale-Bopp    &  0.01 -- 0.08 & \protect{\refcite{Crov04}} \\
CH$_3$OD/CH$_3$OH  &   $<0.03$     & Hale-Bopp  & 0.01 -- 0.06  & \protect{\refcite{Crov04}} \\
CH$_2$DOH/CH$_3$OH &  $<0.008  $ & Hale-Bopp    &  0.04 & \protect{\refcite{Crov04}} \\
HDS/H$_2$S &   $<0.2$    & Hale-Bopp    & 0.01 -- 0.1  & \protect{\refcite{Crov04}} \\
\botrule
\end{tabular}
}
\begin{tabnote}
{\small $^a$ Representative interstellar values are from
Ref.~\protect{\refcite{Roueff03}}\@. $^b$ HDO/H$_2$O ratios were
measured in Halley, Hyakutake, \& Hale-Bopp\protect{\cite{Eber95,BoMo98,Meier98a}},
with three further tentative detections in comets Ikeya-Zhang, LINEAR
A2, and Lee\protect{\cite{Gibb02}} \\}
\end{tabnote}
\label{sbc:tab:dh}
\end{table}

With the exception of the large HDCO/H$_2$CO measured by Kuan {\it et
al.}\cite{Kuan07}, cometary D/H ratios are somewhat smaller than
those measured in dark interstellar clouds and low-mass star-forming
regions (hot corinos)\cite{CeccPPV,Helen06}, but are in line with
those found in massive hot cores\cite{Jacq90}. The HDO and DCN
fractionation in the ISM is understood to result from ion-molecule
isotope-exchange reactions at 10~K, whereas the cometary
values are more in line with a somewhat warmer temperature of $\approx
30$~K\cite{MBH89}\@.

\subsection{Nitrogen}

$^{15}$N/$^{14}$N ratios have only been observed in two cometary
molecules: hydrogen cyanide and the related cyanogen radical. 
HC$^{15}$N/HC$^{14}$N ratios of 0.003 were seen in
Hale-Bopp\cite{Jewitt97,Ziurys99}, roughly equal to the terrestrial
nitrogen isotope ratio, and slightly larger than the
elemental protosolar ratio\cite{Owen01}. C$^{15}$N/C$^{14}$N ratios
have been measured in eight comets, including Hale-Bopp, with an
apparently constant value of 0.007 in every comet\cite{Huts05}. The
discrepancy between the isotope ratios in HCN and CN shows that HCN
cannot be the sole parent of CN, and demonstrates the importance of
isotope ratios in testing putative `parent-daughter' relationships
between coma species\cite{RC02PSSb}. The CN is thought to originate
from the break-up of $^{15}$N-enriched organic refractory (CHON) material in
the coma\cite{Huts05}. Interestingly, $^{15}$N-rich organics have been
discovered in {\it Stardust} samples\cite{Sand06,McK06}{}, and also in
meteorites and IDPs\cite{Buse06,Floss06}\@.

Cometary (and meteoritic) $^{15}$N fractionation is often assumed to
originate in the ISM via low-temperature chemistry, in analogy with
enhanced D/H ratios, as there are no known nebular processes that can
produce such effects.  However, there have been very few observational
studies of nitrogen fractionation in the ISM\@. Theoretical
calculations
predicted only small
$^{15}$N-enhancements of $\sim 25$ per cent in a `typical' dark
interstellar cloud\cite{TH00}. More recent studies
indicate that, if CO is depleted from the gas-phase, much larger
enhancements are possible\cite{N1502,N1504}{}. In this scenario,
$^{15}$N is preferentially incorporated into ammonia ice, and bulk ice
$^{15}$N-enhancements of 80 per cent are possible. Following the
isotopic ratios in individual monolayers (ML) as they accrete
sequentially, these models also show that most of the
$^{15}$N-enhancement is due to highly-fractionated uppermost ML which
accrete at late times\cite{N1507}. These layers are also the most
likely to be altered by subsequent processing of the ices by UV and
cosmic ray irradiation into more refractory material, although the
details of this processing are unclear. If this mechanism is in fact
the origin of the $^{15}$N anomalies in primitive solar system
material, this indicates that at least some cometary organics
originally formed in cold (10~K) interstellar gas.

\subsection{Oxygen and carbon}
  
Gas-phase $^{13}$C/$^{12}$C ratios have been measured in cometary C$_2$, CN, and
HCN\@. To date, the only $^{18}$O-bearing isotopologue
detected in a comet is H$_2^{18}$O\@. For both elements, the isotopic
ratios are all `normal' (i.e., Solar)\cite{BoMo04}, although the error
bars in each case are sufficiently large that small deviations from
the Solar ratio would not have been identified.
More accurate laboratory analysis of the {\it Stardust} samples did in
fact reveal the presence of some cometary grains with small $^{18}$O and $^{17}$O
depletions, of the order forty parts per mil\cite{McK06}. This
material is isotopically identical to many particles found in
meteorites and IDPs\cite{Lyons05}, and presumably shares a common
origin. Similarly small $^{13}$C-anomalies are present in many {\it
Stardust} particles, also in line with those observed in other
primitive solar system material\cite{ClayARAA04}. Some of these phases
are apparently `pre-solar' grains preserved intact since their
formation in the outflows of red giant stars or supernovae (e.g.,
Ref.~\refcite{Lodders05}), but in other cases the fractionation may be
due to chemistry in the ISM\@.

Low temperature interstellar fractionation of oxygen and carbon via
ion-molecule reactions was modeled by Langer et
al.\cite{Langer84}, who showed that relatively small effects are to
be expected. For carbon, $^{13}$C becomes preferentially incorporated
into CO, whereas other C-bearing species become
$^{13}$C-deficient. Subsequent freeze-out and surface chemistry will
preserve these differences in specific molecules\cite{SBCPIL}. For
example, formaldehyde and methanol are thought to be formed via
hydrogenation of CO, so one would expect these species to have a
slight $^{13}$C excess.
Ion-molecule reactions are apparently unable to account for the
observed oxygen isotope anomalies, which instead have been interpreted
as arising from self-shielding of the more abundant $^{12}$C$^{16}$O
molecule compared to its isotopologues\cite{Lyons05,Thiemens83}. This
shielding may have occurred at the surface of the interstellar cloud
from which the solar system was formed\cite{Yuri04}, or at the surface
of the protosolar nebula\cite{Lyons05}. In either case, the
fractionation occurred in UV-irradiated gas, which models of
photon-dominated regions (PDRs) predict to have temperatures of $\sim
30$--70~K\cite{Tielens85}\@. Thus, the presence of oxygen isotope
anomalies in cometary material is evidence for the formation of some
cometary matter in warm gas.

\section{From the pre-Solar core to comets: tracing the chemical
  heritage of cometary material}
\label{sec:cocoon} 

\begin{figure*}[t]    
\begin{center} 
\includegraphics[width=4.6in]{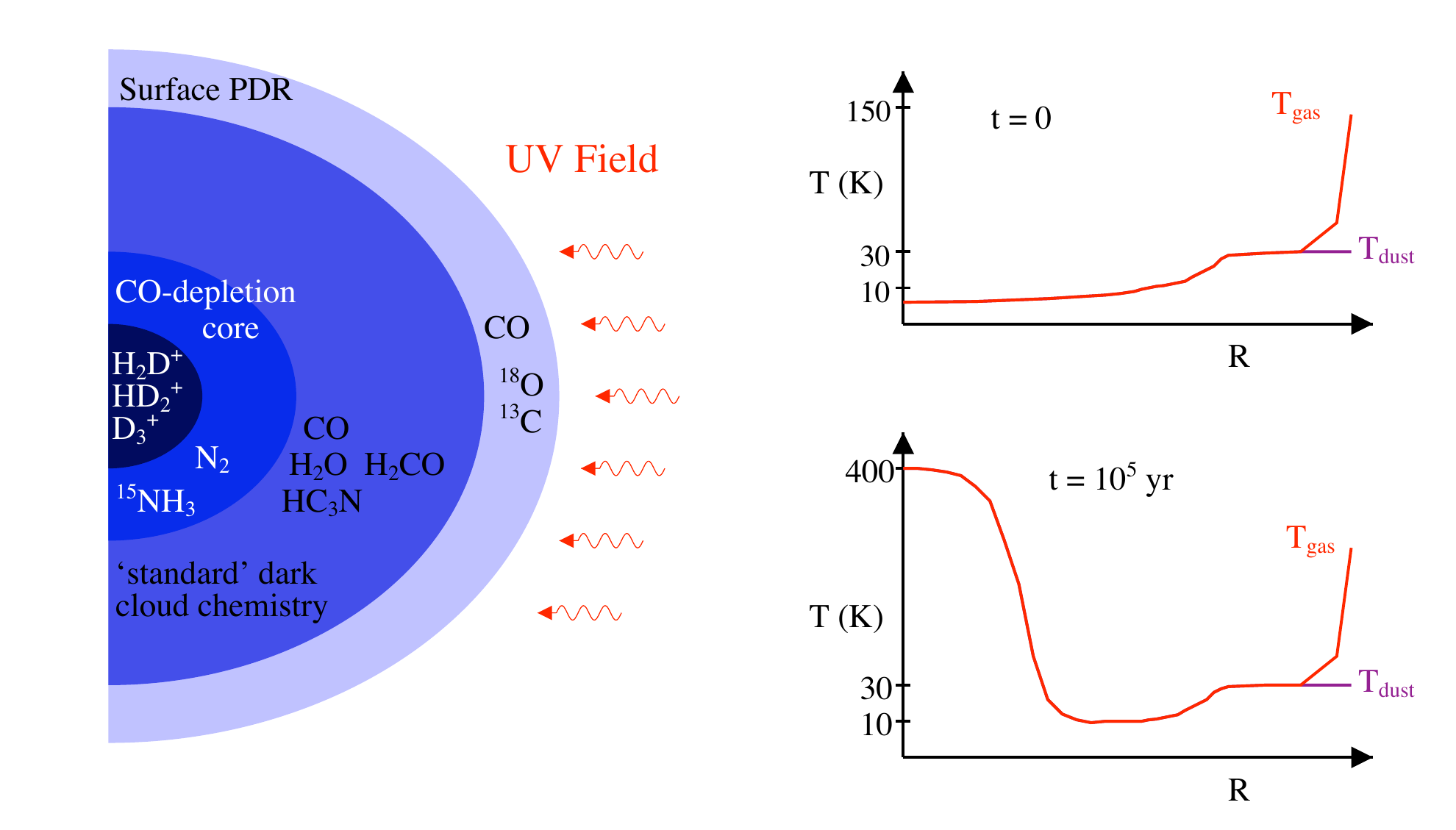}  
\caption{Schematic view of the chemical and thermal structure of a
  pre-stellar and proto-stellar core (not to scale).}
\label{sbc:fig:corecartoon}
\end{center}  
\end{figure*}

In the previous section we described the isotopic ratios seen in
cometary material, and the possible relations between cometary and
interstellar material. In this section, we briefly describe a model of
core collapse and isotope fractionation that can plausibly account for the
observational data. \Fref{sbc:fig:corecartoon} illustrates the
chemical structure of a pre-stellar core, immediately prior to the
onset of collapse and beginning of the Class 0 phase. The core
contains strong chemical gradients, and we expect these gradients will
result in related chemical differentiation (both spatial and temporal)
during  the subsequent gravitational collapse.

In the coldest, most dense,  regions toward the center all heavy elements
are frozen on to grains. Here, D$_3^+$ dominates the
gas-phase\cite{Roberts03}, and accretion of gas with  high atomic D/H ratios  leads to
ices containing   highly-deuterated molecules\cite{SBCetal97}{}.
However, once the core begins to collapse, most of the material in
this centermost region will end up in the newborn protostar. 
Hence, we may expect the most heavily-deuterated
molecules detected in the ISM to generally be only a minor constituent of
comets. Surrounding this region, a shell of CO-depleted gas is the site
of efficient $^{15}$N-fractionation, which produces $^{15}$N-enriched
ammonia ice after about  a few times
$10^5$~yr\cite{N1502,N1504}{} - this  time-scale is of sufficient duration that, for the  collapse of a 1~$M_\odot$ core,   this shell will collapse  during   early Class I phase.

Farther out, CO remains in the gas phase and the chemical composition
can be understood from `standard' models of dark cloud chemistry.
This region will however contain a temperature gradient, with those shells
immediately behind the PDR  being warmed by infrared reradiation of the UV
absorbed at the surface,  and so having gas and dust temperatures   of $\approx
25$--35~K\@.  As these regions are almost the last to accrete, during the  Class I/ Class
II epochs, they are a major source of material in the disk at the
time when comets were formed and will be characterised by reduced deuterium
fractionation and molecular  ortho:para spin 
ratios characteristic of these temperatures \cite{BoMo04}\@

The interstellar UV field to which the surface of the core is exposed may actually be much
stronger than the mean interstellar flux if the core is in close
proximity to newborn massive O and B stars\cite{Adams04}, and so
warm gas and dust may exist to greater  depths.  
 In any case, self-shielding of CO  in these surface layers
leads to mass-independent fractionation of atomic $^{13}$C,
$^{17}$O, and $^{18}$O\cite{Yuri04,Lyons05}. The UV flux also results
in increased gas and dust  temperatures interior to the surface PDR, with $T_{\rm gas} \sim 70$--150~K and $T_{\rm dust} \sim 30$--35~K\cite{Tielens85}\@.  
   Previous models of CO self-shielding in clouds and disks have relied on rapid transport  (e.g. turbulent diffusion) from the PDR to cold regions, where atom sticking and hydrogenation on dust  fixes the oxygen isotopic anomalies in water   \cite{Yuri04,Lyons05,YuriPPV}{}.
 At the  elevated   dust   temperatures in our model (Figure 5), atoms do not stick to the surfaces of dust grains  \cite{SBC97Cores}, so surface chemistry is unable to convert the isotope fractionation in the atoms into fractionation of more refractory species. 
However,  as the gravitational collapse proceeds and the  central luminosity increases,   neutral chemistry in the hot infalling gas provides an alternative gas phase route to O $\rightarrow$ H$_2$O conversion.

\begin{figure*}[t]    
\begin{center} 
\includegraphics[width=4in]{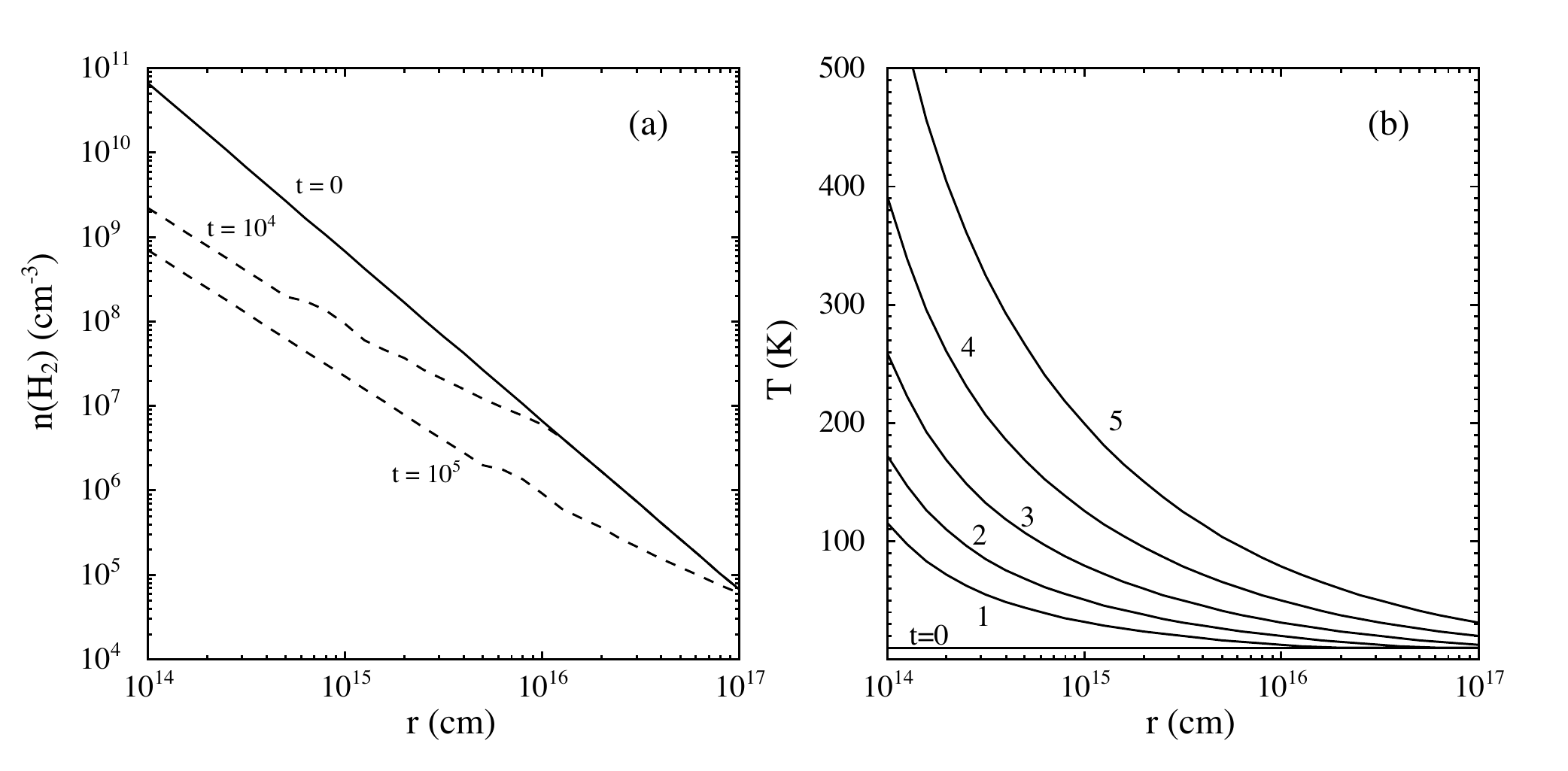} 
\caption{Physical conditions in a collapsing protostellar envelope (see
Ref.~\refcite{SDRCocoon}).}
\label{sbc:fig:cocoon}
\end{center}  
\end{figure*}

Modelling the chemistry in the collapsing protostellar cocoon requires a  physical model in
order to derive density, temperature, and infall velocity profiles in the envelope
as a function of time. We have adapted the  dynamical-chemical model of Rodgers and Charnley\cite{SDRCocoon}{},    based on the  `inside-out' collapse of Shu\cite{Shu77,Adams85}, to include a  static PDR at the outer edge of the envelope. 
This work will be reported in detail  elsewhere\cite{SDR08} and here we present some preliminary results to demonstrate how interstellar fractionation patterns in $^{13}$C, $^{17}$O, and $^{18}$O may be transported from the distant  outer envelope to the central protostar and disk.

\begin{figure*}[t]    
\begin{center} 
\includegraphics[height=3in, width=4in]{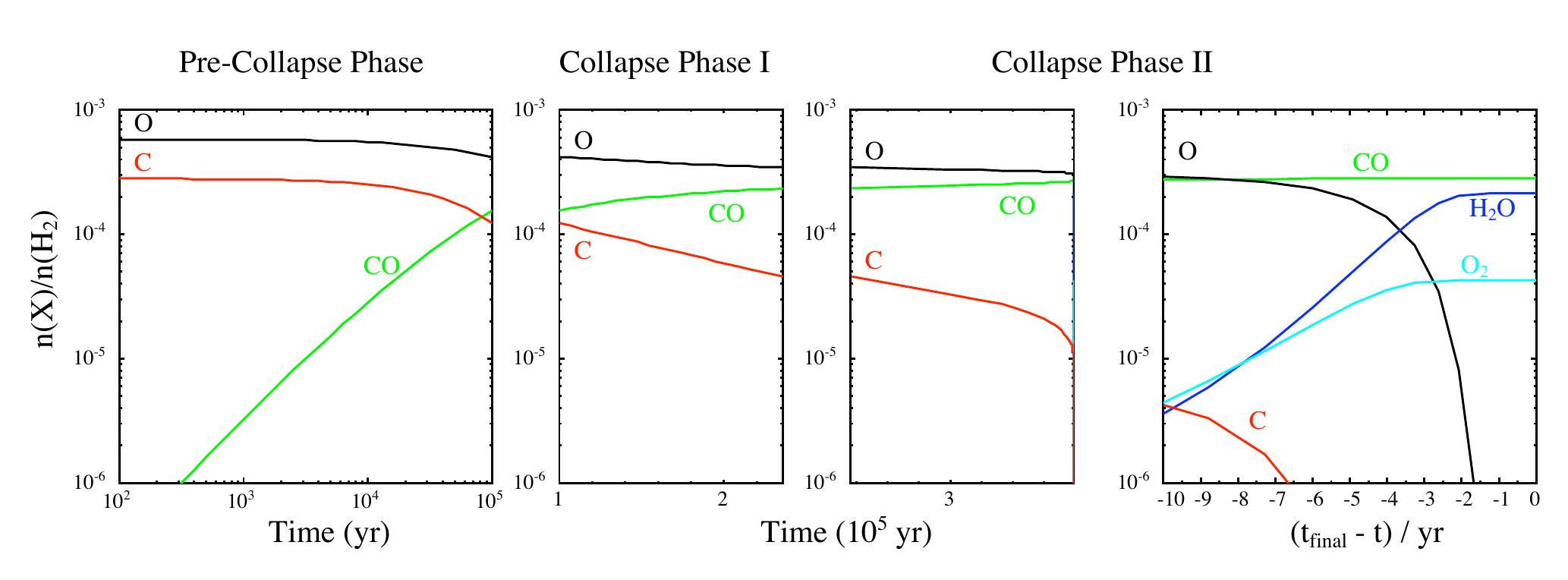} 
\caption{Chemistry in a collapsing protostellar envelope}
\label{sbc:fig:Ochem}
\end{center}  
\end{figure*} 

  \Fref{sbc:fig:cocoon} shows the density and temperature profiles used in
this model, with conditions appropriate for a 1~M$_{\odot}$ star accreting
material at a rate of $10^{-5}$ M$_{\odot}$~yr$^{-1}$. Clearly, at late times when
most of the mass is in the central star, the luminosity is large enough to
significantly heat material in the innermost regions of the collapsing envelope.
However, by this point the infall velocities are sufficiently large to ensure that
the dynamical time-scale for material to pass through this hot zone is much less
than the typical time-scales for chemical reactions to alter the composition of
the gas\cite{SDRCocoon}{}. Important exceptions are the reactions of O and OH with
H$_2$ which rapidly convert atomic oxygen into water.  

This is illustrated
in \fref{sbc:fig:Ochem}, which shows the C and O chemistry in material initially
located at a radius of $10^{4}$~AU\@. The far right panel is a blow-up of the final ten years,
showing how high-$T$ chemistry drives O into H$_2$O near the protostar. 
Thus, the  material that  rains down onto the  disk,  during the epoch  that cometesimals and planetesimals are forming,  contains  isotopically-enhanced
water derived from  the isotopic anomalies in neutral oxygen atoms generated by CO self-shielding  in the surface PDR.


\section{Conclusions}
\label{sec:conc}

We have reviewed the putative contribution of  interstellar chemistry to the volatile composition of comets.  In doing so, we have purposely neglected discussion of the possible contribution of nebular chemistry as a source of  cometary volatiles  - either from the warm innner nebula or from an essential continuation of interstellar chemistry in the cold outer disk; these have been reviewed elsewhere\cite{AJM04,Ciesla06}.

Comets could contain material from  several different stages of molecular  cloud evolution and we have illustrated the related chemistry with some recent observations. 
 Many of the molecules detected in comets, including several  organics, are either directly observed in interstellar ices (e.g. methanol, formic acid) or are believed to form on interstellar grains (e.g. formamide)\cite{ARAA00}.  A direct comparison of the organic inventories is however complicated by the fact that, in comets, additional sources of simple molecules appear  to contribute to their coma abundances (e.g. CO, CS, CN, formaldehyde). These so-called {\it extended sources} are believed to be due to the thermal or photolytic break-up of large organic macromolecules\cite{Cottin07} and have no readily identifiable parallel in interstellar chemistry.     
 Interstellar isotopic fractionation in dense gas with temperatures in the range $\sim 10$--35~K  could account for the currently known, albeit meagre,  fractionation  ratios measured in comets.   
 
 We have presented the outline of a model that may account for all these characteristics as arising in the prestellar core from which the Sun formed.  As observed in many galactic sources, such cores can develop very strong chemical gradients. Gravitational collapse of such a core may deliver chemically distinct  regions (i.e.\ mass shells)  onto the central disk during the Class 0/I epoch, and later, to the extent that all the `interstellar' characteristics of comets are delivered during this infall. 
   
Finally,  future observations of more bright comets, especially short-period ones,  as well as  of protoplanetary disks, are necessary (e.g.\ \refcite{Lahuis 06}).   The advent of the {\it Atacama Large Millimetre Array} promises a great advances in further investigating the ISM-comet
connection\cite{vanD07,WilsonALMA}.

\section*{Acknowledgements}

This work was  supported by NASA's Origins of Solar Systems, Planetary Atmospheres and Exobiology  Programs and by the NASA Goddard Center for Astrobiology. We are grateful to Shigehisa Takakuwa for preparing Figure 3.  A.J. Markwick made the map of Figure 2.



\begin{thebibliography}{96}



\bibitem{Pascale04}Ehrenfreund, P.,  Charnley, S.~B., \& Wooden, D.\ 2004,
in Comets II, eds.\ M. C. Festou, H. U. Keller, \& H. A. Weaver, 
University of Arizona Press, Tucson, p.~115
 
\bibitem{ARAA00}Ehrenfreund, P., Charnley S.B., 2000, Annu.\ Rev.\ A \& A, 38, 427

\bibitem{Brownlee06} Brownlee D., et al., 2006, Science, 314, 1711

\bibitem{Wooden02} Wooden D.H. 2002, EM\&P, 89, 247

\bibitem{Wooden07} Wooden D.H., Desch S., Harker D., Gail H.-P.,
  Keller L, 2007,  in Protostars and Planets V, eds.\
B. Reipurth, D. Jewitt, \& K. Keil, University of Arizona Press, Tucson, p.~815

\bibitem{PPVBook} Reipurth B., Jewitt D., Keil K., 2007, (eds.), Protostars and
  Planets V, University of Arizona Press, Tucson

\bibitem{Lada07} Lada C.J., Alves J.F., Lombardi M., 2007, in
Protostars and Planets V, eds.\
B. Reipurth, D. Jewitt, \& K. Keil, University of Arizona Press, Tucson, p.~3

\bibitem{BerginARAA07} Bergin E.A., Tafalla M., 2007, Annu.\ Rev.\ A \& A, 45, 339

\bibitem{vanD07} van Dishoeck E.F., J{\o}rgensen, J.K., 2007, Ap\&SS, 337, 15

\bibitem{Hueso05} Hueso R., Guillot T., 2005, A\&A, 442, 703

\bibitem{AJM04} Markwick, A. J., Charnley, S. B. 2004, in 
Astrobiology: Future Perspectives, eds.\ P.Ehrenfreund et al., 
Kluwer, Dordrecht, p.~33

\bibitem{Thi04} Thi, W.-F., van Zadelhoff, G.-J., van Dishoeck, E.F., 2004, A\&A, 425, 955

\bibitem{vanD06} van Dishoeck E.F., 2006, Proc.\ Nat.\ Acad.\ Sci., 103, 12249

\bibitem{diFrancescoPPV} di Francesco, J., 
Evans, N.~J., II, Caselli, P., Myers, P.~C., Shirley, Y., Aikawa, Y., \& 
Tafalla, M.\ 2007, in Protostars and Planets V, eds.\
B. Reipurth, D. Jewitt, \& K. Keil, University of Arizona Press, Tucson, p.~17 

\bibitem{CeccPPV} Ceccarelli, C., 
Caselli, P., Herbst, E., Tielens, A.~G.~G.~M., \& Caux, E.\ 2007,
in Protostars and Planets V, eds.\
B. Reipurth, D. Jewitt, \& K. Keil, University of Arizona Press, Tucson, p.~47

\bibitem{BerginPPV} Bergin, E.~A., Aikawa, Y., Blake, G.~A., \& van Dishoeck, E.~F.\ 2007,
in Protostars and Planets V, eds.\
B. Reipurth, D. Jewitt, \& K. Keil, University of Arizona Press, Tucson, p.~751 

\bibitem{vanDBlake98} van Dishoeck E.F., Blake G.A., 1998, Annu.\ Rev.\ A \& A, 36, 317

\bibitem{Mundy00} Mundy L.G., Looney L.W., Welch W.J., 2000,
in Protostars and Planets IV, eds.\ V. Mannings, A.P. Boss \& S.S. Russell,
U. Arizona Press, p.~355

\bibitem{Evans06} Evans N.J., et al., 2006, Bull.\ AAS, 38, 1204

\bibitem{Ceccarelli06} Ceccarelli C., 2006, in Astrochemistry: Recent
  Successes and Current Challenges, eds.\ D.C. Lis et al., Cambride
  Universtiy Press, p.~1

\bibitem{SDRCocoon} Rodgers S.D., Charnley S.B., 2003, ApJ, 585, 355

\bibitem{Lee05} Lee J-E., Evans N.J., Bergin E.A., 2005, ApJ, 631, 351

\bibitem{Cassen81} Cassen P., Moosman A., 1981, Icarus, 48, 353

\bibitem{Butner07} Butner H.M., et al., 2008, in preparation

\bibitem{Goldsmith86} Goldsmith, P.F., Langer W.D., Wilson R.W., 1986, ApJ, 303, L11

\bibitem{Fuller91} Fuller G.,A., et al., 1991, ApJ, 376, 135

\bibitem{Beichman84} Beichman C.A., et al., 1984, ApJ, 278, L45

\bibitem{Yu99} Yu, K.C., Billawala, Y., Bally, J., 1999, AJ, 118, 2940

\bibitem{Langer96} Langer W.D., et al., 1996, ApJ, 468, L41

\bibitem{Buckle06}Buckle, J. V., Rodgers, S. D., Wirstrom, E. S., Charnley,
 S. B.,  Markwick-Kemper, A. J., Butner, H. M., Takakuwa S., 2006,
 Faraday Discussions, 133, 63

\bibitem{Charnleyetal07} Charnley S.B., et al., 2008,  in preparation

\bibitem{Olano88} Olano C.A., Walmsley C.M., Wilson T.L., 1988, A\&A, 196, 194

\bibitem{Hirahara92} Hirahara Y., et al., 1992, ApJ, 394, 539

\bibitem{HerbstLeung89} Herbst E., Leung, C.M., 1989, ApJS, 69, 271

\bibitem{Peng98} Peng R., et al., 1998, ApJ, 497, 842

\bibitem{Ohishi98} Ohishi M., Kaifu N., 1998, Faraday Discussions, 109, 205

\bibitem{AJM2000} Markwick A.J., Millar T.J., Charnley S.B., 2000,
  ApJ, 535, 256

\bibitem{Caselli02} Caselli P., Benson P.J., Myers P.C.,  Tafalla M., 2002, ApJ, 572, 238

\bibitem{Hirota02}Hirota T., Ito T., Yamamoto S., 2002, ApJ, 565, 359

\bibitem{Hirota04}Hirota T., Maezawa H., Yamamoto S., 2002, ApJ, 617, 399

\bibitem{Bergin01} Bergin E.A., et al., 2001, ApJ, 557, 209

\bibitem{Bergin02} Bergin E.A., Alves J., Huard T., Lada C.J., 2002, ApJ, 570, L101

\bibitem{Tafalla02} Tafalla M., Myers P.C., Caselli P., Walmsley C.M.,
  Comito C., 2002, ApJ, 569, 815

\bibitem{SBC94} Charnley S.B., 1994, in Molecules and Grains in Space,
  ed.\ I. Nenner, AIP Press, New York, p.~155

\bibitem{SBC97Cores} Charnley S.B., 1997, MNRAS, 291, 455

\bibitem{Bergin97} Bergin E.A., Langer W.D., 1997, ApJ, 486, 316

\bibitem{Wooden04}  Wooden, D. H., Charnley, S. B., Ehrenfreund, P. 2004,
in Comets II, eds.\ M.C. Festou, H.U. Keller \& H.A. Weaver,
U. Arizona Press, Tucson, p.~33

\bibitem{Cazeaux03} Cazaux S., Tielens A.G.G.M., Ceccarelli C.,
  Castets A., Wakelam V., Caux E., Parise B., Teyssier D., 2003, ApJ, 593, L51

\bibitem{Kuan04} Kuan, Y.-J., et al., 2004, ApJ, 616, L27

\bibitem{Bottinelli04b} Bottinelli S. et al., 2004, ApJ, 615, 354

\bibitem{Bottinelli04a} Bottinelli S. et al., 2004, ApJ, 617, L69

\bibitem{Bottinelli07} Bottinelli S., 
Ceccarelli C., Williams J.P., Lefloch B., 2007, A\&A, 463, 601

\bibitem{Jorgensen05} J{\o}rgensen J.K., Bourke T.L., Myers P.C.,
  Sch{\"o}ier F.L., van Dishoeck E.F., Wilner D.J., 2005, ApJ, 632, 973

\bibitem{Sakai06} Sakai, N., Sakai, T., Yamamoto, S.\ 2006, PASJ, 58, L15



\bibitem{Mayo} Greenberg J.M., 1982, in COMETS, ed.\ L.L. Wilkening,
U. Arizona Press, p.~131

\bibitem{IrvPPIV}Irvine, W.~M., Schloerb, 
F.~P., Crovisier, J., Fegley, B., Jr., Mumma, M.~J.\ 2000.\
in Protostars and Planets IV, eds.\ V. Mannings, A.P. Boss \& S.S. Russell,
U. Arizona Press, p.~1159

\bibitem{ISSI} Charnley S.B., Rodgers S.D. 2008,
Space Sci.\ Rev., in press

\bibitem{RC02} Rodgers, S.~D., Charnley, S.~B., 2002, MNRAS, 330, 660 

\bibitem{Roueff03}Roueff E., Gerin M., 2003. Space Sci.\ Rev., 106, 61

\bibitem{Sand06}Sandford S., et al., 2006, Science, 314, 1720

\bibitem{Mess00}Messenger S., 2000, Nature, 404, 968

\bibitem{Meier98b} Meier R., et al., 1998, Science, 279, 1707 

\bibitem{Kuan07} Kuan Y-J. et al., 2008, {in preparation}  

\bibitem{Kawa05} Kawakita H., et al., 2005, ApJ, 623, L49

\bibitem{Crov04} Crovisier J. et al., 2004a, A\&A, 418, 1141

\bibitem{Eber95} Eberhardt, P., Reber, 
M., Krankowsky, D., Hodges, R.~R.\ 1995.
A\&A, 302, 301

\bibitem{BoMo98} Bockel\'ee-Morvan D., et al.\ 1998, Icarus, 133, 147

\bibitem{Meier98a} Meier R., et al., 1998, Science, 279, 842  

\bibitem{Gibb02} Gibb, E.~L., Mumma, M.~J., 
Disanti, M.~A., dello Russo, N., Magee-Sauer, K.\ 2002.\
in Asteroids, Comets, and Meteors 2002, ed.\ B. Warmbein, ESA, p.~705 

\bibitem{Helen06} Roberts H., 2006, in Astrochemistry: Recent
  Successes and Current Challenges, eds.\ D.C. Lis et al., Cambride
  Universtiy Press, p.~27

\bibitem{Jacq90}Jacq T., et al., 1990, A\&A, 271, 276

\bibitem{MBH89} Millar T.J., Bennett A., Herbst E., 1989, ApJ, 340, 906

\bibitem{Jewitt97} Jewitt, D., Matthews, 
H.~E., Owen, T., Meier, R., 1997,
Science, 278, 90 

\bibitem{Ziurys99} Ziurys, L.~M., Savage, 
C., Brewster, M.~A., Apponi, A.~J., Pesch, T.~C., Wyckoff, S., 1999.,
ApJ, 527, L67 

\bibitem{Owen01} Owen, T., Mahaffy, P.~R., 
Niemann, H.~B., Atreya, S., Wong, M.\ 2001, 
ApJ, 553, L77 

\bibitem{Huts05}
Hutsem{\'e}kers, D., Manfroid, J., Jehin, E., Arpigny, C., Cochran, A., 
Schulz, R., St{\"u}we, J.~A., Zucconi, J.-M., 2005,
A\&A, 440, L21 

\bibitem{RC02PSSb}Rodgers, S.~D., Charnley, S.~B., 2002,
P\&SS, 50, 1215 

\bibitem{McK06}McKeegan K. D. et al., 2006, Sci., 314, 1724

\bibitem{Buse06} Busemann H., Young A.~F., Alexander C.~M.~O'D., Hoppe P.,
Mukhopadhyay S., Nittler L.~R., 2006, Science, 312, 727

\bibitem{Floss06}Floss C., Stadermann  F.~J., Bradley J.~P., Dai Z.~R., 
Bajt S., Graham G., Lea A.~S., 2006, Geo.\ Cosmo.\ Acta, 70, 2371

\bibitem{TH00}Terzieva R., Herbst E., 2000, MNRAS, 317, 563

\bibitem{N1502}Charnley S.B., Rodgers S.D., 2002, ApJ, 569, L133

\bibitem{N1504}Rodgers S. D., Charnley S. B., 2004, MNRAS, 352, 600

\bibitem{N1507}Rodgers S. D., Charnley S. B., 2008, MNRAS, 385, L48

\bibitem{BoMo04} Bockel\'ee-Morvan D., et al., 2004,
in Comets II, eds.\ M.C. Festou, H.U. Keller \& H.A. Weaver,
U. Arizona Press, Tucson, p.~391

\bibitem{Lyons05}Lyons J.R., Young E.D., 2005, Nature, 435, 317

\bibitem{ClayARAA04} Clayton D.~D., Nittler L.~R., 2004, ARA\&A, 42, 39

\bibitem{Lodders05}Lodders K., Amari S., 2005, Chemie der Erde, 65, 93

\bibitem{Langer84} Langer W.D., et al., 1984, ApJ, 277, 581

\bibitem{SBCPIL} Charnley S.B., et al., 2004, MNRAS, 347, 157

\bibitem{Thiemens83} Thiemens M., Heidenreich J.E., 1983, Science,
  219, 1073

\bibitem{Yuri04} Yurimoto H., Kuramoto K., 2004, Science, 305, 1763

\bibitem{Tielens85}Tielens A.G.G.M., Hollenbach D.J., 1985, ApJ, 291, 722

\bibitem{Roberts03}Roberts, H., Herbst, E., Millar, T.J. 2003, ApJ, 591, L41

\bibitem{SBCetal97}Charnley, S.B., Tielens, A.G.G.M., Rodgers, S.D. 1997, ApJ, 482, L203 

\bibitem{Adams04} Adams, F.C. et al., 2004, ApJ, 611, 360

\bibitem{YuriPPV} Yurimoto H., et al., 2007,  in Protostars and Planets V, eds.\
B. Reipurth, D. Jewitt, \& K. Keil, U. Arizona Press, Tucson, p.~849

\bibitem{Shu77}Shu F., 1977, ApJ, 214, 488

\bibitem{Adams85}Adams F.C., Shu F.H., 1985, ApJ, 296, 655

\bibitem{SDR08} Rodgers S.D., Charnley S.B., 2008, in preparation

\bibitem{Ciesla06} Ciesla F.J., Charnley S.B., 2006,
in Meteorites and the Early Solar System II, eds.\ D. S. Lauretta \& H. Y. McSween Jr.,
U. Arizona Press, Tucson, p.~209

\bibitem{Cottin07}Cottin H., Fray N., 2007, Space Sci.\ Rev., in press

\bibitem{Lahuis06} Lahuis F., et al., 2006, ApJ, 636, L145


\bibitem{WilsonALMA} Wilson A., 2005, (ed.), The dusty and molecular universe:
a prelude to Herschel and ALMA (ESA SP-577), ESA Publications, Noordwijk.





%
%
%

\end{thebibliography}
\end{document}